\documentclass[11pt,twoside]{article}

\usepackage{asp2006}
\usepackage{epsf}
\usepackage{graphicx}
\usepackage{lscape}

\begin{document}
\title{The High-Density Ionized Gas in the Central Parsecs of the Galaxy}   
\author{Jun-Hui Zhao, Ray Blundell, James M. Moran}   
\affil{Harvard-Smithsonian Center for Astrophysics, 60
Garden Street, MS 78, Cambridge, MA 02138}    
\author{D. Downes, Karl F. Schuster}
\affil{Institut de Radio Astronomie Millim\'etrique, 38406 Saint Martin d'H\`eres, France}
\author{Dan Marrone}
\affil{Department of Astronomy, University of Chicago, Chicago, IL 60637}
\begin{abstract} 
We report the results from observations of  H30$\alpha$ line emission 
in Sgr A West with the Submillimeter Array at a resolution of 2\arcsec\
and a field of view of about 40\arcsec. The H30$\alpha$ line is sensitive 
to the high-density ionized gas in the minispiral structure. We compare 
the velocity field obtained from H30$\alpha$ line emission to a Keplerian 
model, and our results suggest that the supermassive black hole at Sgr A* 
dominates the dynamics of the ionized gas. However, we also detect 
significant deviations from the Keplerian motion, which show that the 
impact of  strong stellar winds from the massive stars along the ionized 
flows and the interaction between Northern and Eastern arms play 
significant roles in the local gas dynamics.

\end{abstract}

\section{Introduction}
The Galactic Center harbors a supermassive black hole (SMBH)
with a mass of $4.2\times10^6$ M$_\odot$ at the position of 
Sgr A* \citep{ghez08,gill09}. Previous radio recombination 
line studies \citep{schw89,rob93,rob96} revealed complex 
structures of ionized gas around Sgr A*, including the ``minispiral'' 
and ``Bar.'' Parts of these structures have been modeled as 
gas orbiting Sgr A* \citep{schw89,sand98,paum04,voll00,lisz03,zhao09} 
while other parts of the structures may be gravitationally unbound 
\citep{yusef98}. In particular, the strong stellar winds from 
the clusters of massive stars in the vicinity of Sgr A*, with 
the help of gravitational focusing, create large-scale, unbound 
features \citep{lutz93}. There are a number of difficulties in 
existing line observations of these structures. In the cm--radio 
maps from the Very Large Array (VLA), the recombination line strength 
is low, the line-to-continuum ratio is low, the bandwidth coverage 
is limited or irregular, and it is especially difficult to detect 
high-velocity line wings. In the IR, the picture given by the 
emission lines from the ionized gas is distorted, because of 
high obscuration by dust. In this paper, we summarize the main 
results from our observations of the H30$\alpha$ line at 
1.3 mm with the SMA from three array configurations, the combination 
of which provides an angular resolution of 2\arcsec.

\section{Distribution of the High-Density Ionized Gas}
Figure 1 presents the image integrated H30$\alpha$ line emission 
observed in the central 1.5 
parsecs, showing the Northern and Eastern arms of the minispiral. 
The Western Arc does not shown well in our image in the H30$\alpha$ 
line since most of the line emission is below the 4$\sigma$ sensitivity 
cutoff. The H30$\alpha$ line spectra are shown toward selected 
regions with IR sources. A detailed comparison of the H30$\alpha$ 
line spectra and the H92$\alpha$ line spectra observed with the 
VLA yields line ratios of H30$\alpha$/H92$\alpha$ in most of the 
regions that are close to the frequency ratio 
$\nu_{H30\alpha}/\nu_{H92\alpha}\sim28$, suggesting that the 
ionized gas in the central parsecs is optically thin and 
under LTE conditions. In some regions, however (e.g., IRS 1W 
and 10W), the line ratios are significantly smaller than the 
LTE value.

\begin{figure}[!ht]
\begin{center}
\includegraphics[scale = 0.69, angle = -90]{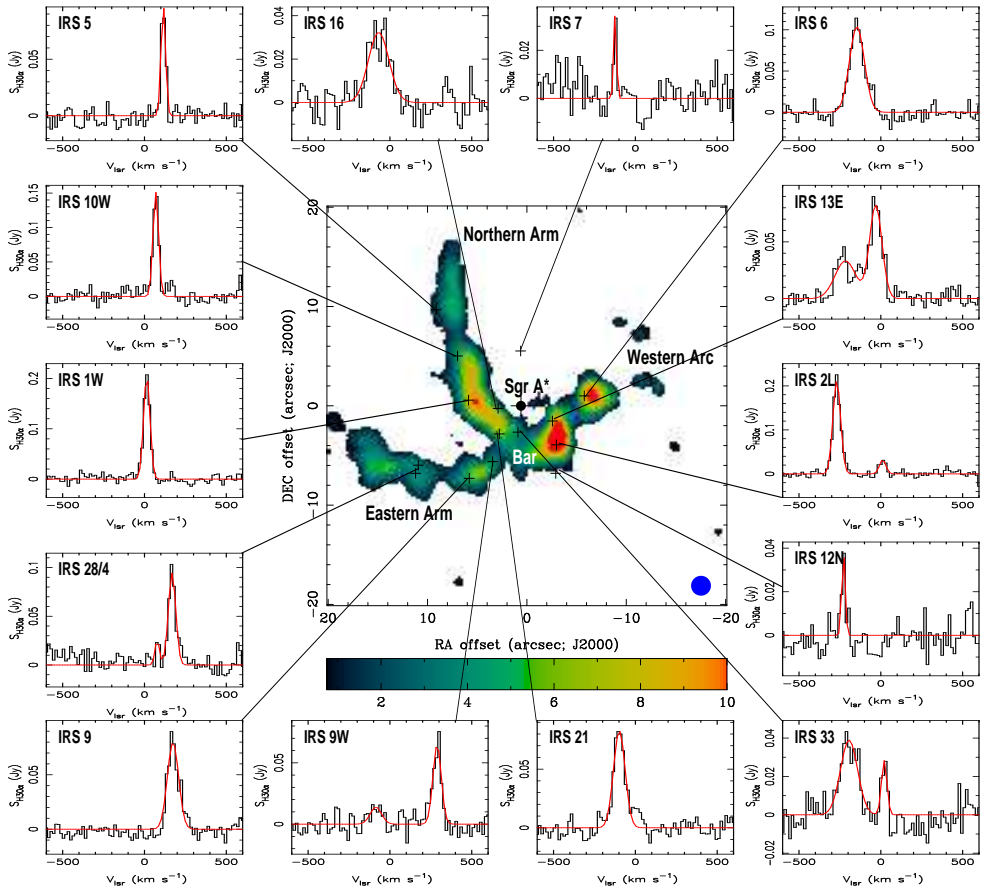}
\end{center}
\caption{The image of the H30$\alpha$ line emission around Sgr A* 
(plus-dot) in the central 1.5-parsec region observed with the SMA
with a resolution of 2\arcsec. The H30$\alpha$ line spectra
of the HII sources within the minispiral are plotted and labelled 
with the names of the corresponding IRS sources.}\label{fig1}
\end{figure}

To assess the physical conditions of the ionized gas, we have also 
fit a model to the observed lines based on the assumption that the 
HII sources are isothermal and homogeneous. The derived typical 
electron densities in the HII clumps are 5--10$\times10^4$ cm$^{-3}$, 
consistent with the values 
($10^4$ cm$^{-3}$) derived from Pa$\alpha$ line emission
at 1.87 $\mu$m \citep{scov03} for  a given  uncertainty
in extinction corrections in the near IR. The typical electron 
temperature in the minispiral arms is 5000--7000 K, in good 
agreement with the previous determination from H92$\alpha$ 
\citep{rob93}. The highest density of $2\times10^5$ cm$^{-3}$ 
occurs in the IRS 13E region while the highest temperatures of 
up to 13000 K were found in the Bar. Both the H30$\alpha$ 
and H92$\alpha$ line profiles are broadened due to the large 
velocity gradients along the line of sight and across the 
2\arcsec\ beam produced by the large dynamical motions near Sgr A*. 
For the H92$\alpha$ line, the line width due to pressure 
broadening appears to be comparable to that of thermal Doppler 
broadening while for the H30$\alpha$ lines, the pressure 
broadening is negligible.

\section{Kinematics and Dynamics}

Figure 2a shows the image of radial velocities determined from fitting the 
peak velocities of the H30$\alpha$ line in the central 40\arcsec, showing 
large velocity gradients along the Northern and Eastern arms. The observed 
radial velocity field is, in general, consistent with that predicted from 
a Keplerian model \citep{zhao09}, suggesting that the SMBH at Sgr A* governs 
the dynamics of the ionized gas. However, in several regions, the observed 
kinematics deviate significantly from those predicted from the Keplerian 
model, as described below.

\begin{figure}[!ht]
\begin{center}
\includegraphics[scale = 0.46, angle = -90]{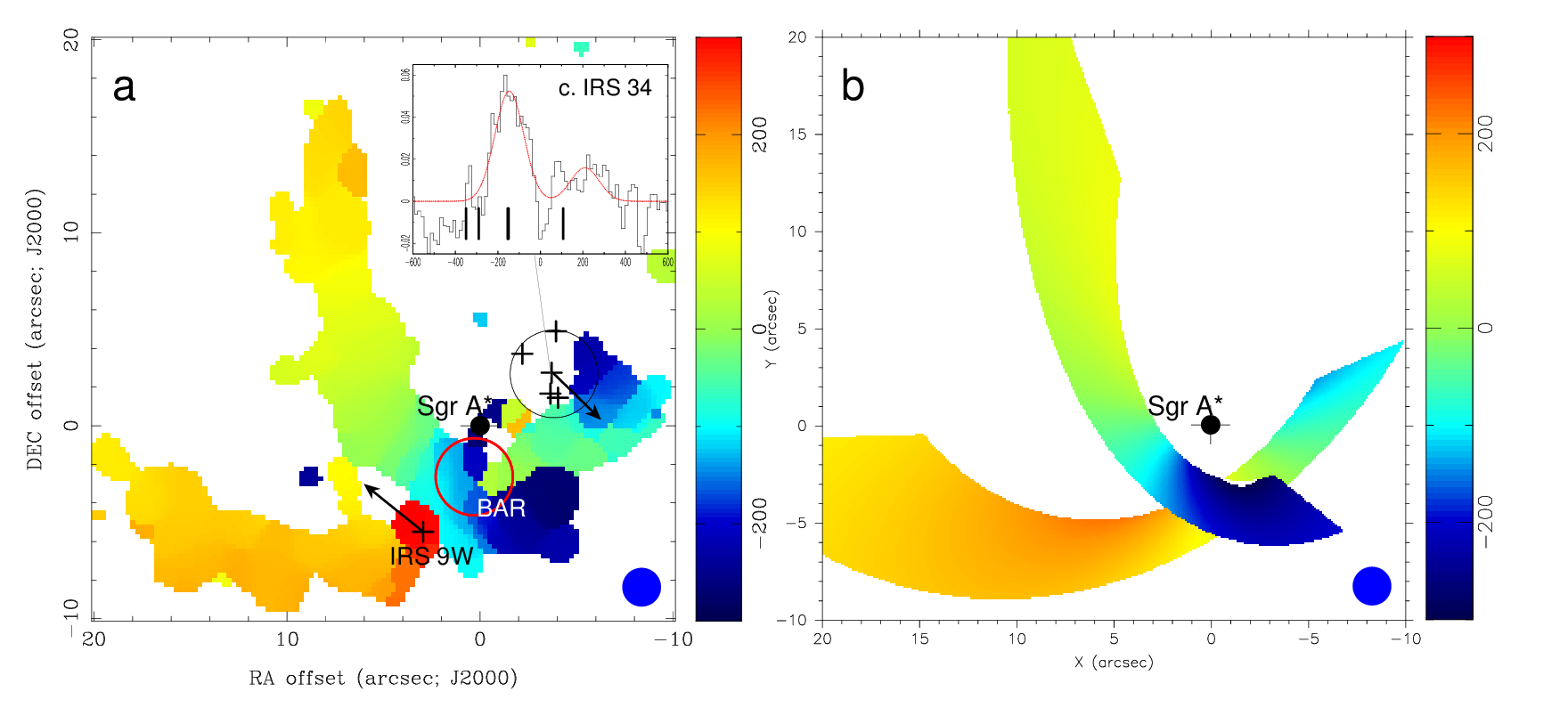}
\end{center}
\caption{The observed radial velocity field ({\bf a}) around Sgr A* (plus-dot)
from the SMA observations of the H30$\alpha$ line is compared with 
that computed from the Keplerian model ({\bf b}). The velocity wedge 
is in units of km s$^{-1}$. The inset ({\bf c}) is the H30$\alpha$ 
spectrum toward the IRS 34 region outlined with the black circle NW 
of Sgr A* in which five massive stars 7SW (WN8), 34W (Ofpe/WN9),
34E (O9-9.5), 34NW (WN7), and 3E (WC5/6) are located (shown with 
``+'' symbols, Paumard et al. 2006). The arrow indicates the direction 
of the averaged transverse velocity of the four stars that have  
proper-motion information. The red circle indicates the region with 
high electron temperature where the interaction between the Northern 
and Eastern arms occurs. The arrow SE of Sgr A* indicates the transverse 
velocity of the IRS 9W star (WN8).
}\label{fig1}
\end{figure}

\noindent {\bf IRS 34:} In the northwest end of the Eastern Arm, a kinematic feature
with radial velocity of up to $-200$ km s$^{-1}$  shows
a large deviation from the value of $-100$ km s$^{-1}$ predicted 
from the Keplerian model (see Figure 2b). Near this region, there is 
a group of Wolf-Rayet (WR) stars, evolved from early-type O stars with 
a mass loss rate of 10$^{-5}$ M$_\odot$ yr$^{-1}$. The suspected 
strong stellar winds from these massive stars may play a 
considerable role in the local kinematics.

\noindent {\bf IRS 9W:} In the ``tip'' region \citep{paum04}, a peculiar 
radial velocity of +300 km s$^{-1}$ was detected from our H30$\alpha$ 
observations, 
showing a large deviation from the radial velocity of +150 km s$^{-1}$
predicted from the Keplerian model. The impact of the strong stellar wind from
the IRS 9W star (WN8)  on the orbiting ionized flow in the Eastern Arm is 
likely responsible for the high velocity of the tip.

\noindent {\bf The Bar:}
Deviations from Keplerian motion are also observed in the Bar region.
These non-Keplerian kinematics could be attributed to the interaction 
between the ionized flows in both Northern and Eastern arms and the 
interaction of ionized flows with the strong winds from the massive 
star clusters, e.g., IRS 16 and 13.

Finally, in the locations with large offsets from Sgr A*, we noticed 
that the deviations in radial velocity from the Keplerian motions become 
significant for the Northern Arm ($>$ 50 km s$^{-1}$) while for the 
Eastern Arm the deviations are less significant. The difference in the 
deviation of the observed radial velocities from the Keplerian velocities 
between the Northern and Eastern arms might suggest that the distribution 
of mass has a preference for the plane of the circumnuclear disk, which is coplanar 
with the orbit of the Northern Arm flow.


\end{document}